\documentclass[%
 reprint,
 amsmath,amssymb,
 aps,
]{revtex4-2}

\usepackage{graphicx}
\usepackage{dcolumn}
\usepackage{bm}
\usepackage{hyperref}

\usepackage{physics}
\usepackage{siunitx}
\usepackage{bbold}

\begin{document}

\preprint{APS/123-QED}

\title{Resonant and collective modification of London dispersion interactions under vibrational strong coupling}

\author{Marit R. Fiechter}
\email{marit.fiechter@phys.chem.ethz.ch}
\affiliation{Department of Chemistry and Applied Biosciences, ETH Z\"urich, 8093 Z\"urich, Switzerland}

\author{Jeremy O. Richardson}
\affiliation{Department of Chemistry and Applied Biosciences, ETH Z\"urich, 8093 Z\"urich, Switzerland}

\date{\today}

\begin{abstract}
Experiments have shown that, by tuning a microcavity to resonance with a vibrational mode of the molecules contained within it, one can modify chemical properties, such as 
reaction rates. This gives rise to the exciting prospect of steering chemical reactivity, just by placing a pair of carefully spaced mirrors around the reaction mixture. 
However, a decade after the first demonstration, the mechanism behind this effect remains ill-understood. 
Here, we show how vibrational strong coupling can lead to resonant modification of vibrationally-resolved London dispersion interactions. Employing a mixed quantum--classical dynamics scheme, we then show how this in turn can give rise to resonant rate enhancement in the case of two molecules strongly coupled to the cavity mode, for all regimes of solvent friction. 
The resonant changes of the London dispersion interaction seem to persist when increasing the number of molecules. 
Whether this also leads to altered reaction rates in the experimentally relevant collective limit remains an open question, as this regime falls outside the range of applicability of our mixed quantum--classical dynamics approach. Nevertheless, the framework presented here offers an exciting new avenue to explore, and hopefully bring us a step closer towards explaining the mechanism behind vibropolaritonic chemistry.

\end{abstract}

\maketitle

\section{Introduction}
10 years ago, it was demonstrated for the first time that strongly coupling a Fabry--P\'erot cavity mode to a molecular vibration (\emph{i.e.} vibrational strong coupling, or VSC) can alter ground-state chemical reactivity \cite{thomas2016ground}. This has spurred a lot of research, both experimental and theoretical, as reviewed in \emph{e.g.} Refs.~\cite{nagarajan2021chemistry,simpkins2021mode,dunkelberger2022vibration,xiang2024molecular,hirai2025polaritonbook,ribeiro2018polariton,fregoni2022theoretical,li2022review,mandal2023theoretical,campos2023swinging,ruggenthaler2023understanding,ying2025collective}. However, on the experimental side open questions remain, amongst other things about the relevance of optical artifacts when spectroscopically tracking the progress of a reaction in a cavity \cite{ simpkins2021mode,thomas2024strong,michon2024impact,michon2025book,vergauwe2026toward}. 
Moreover, in spite of a decade of theoretical work, no conclusive explanation for the mechanism behind these changes in chemical reaction rates has been identified yet.

For a theory to explain experiment, there are two main criteria it has to fulfil. Firstly, the effect a theory predicts has to be \emph{resonant}, so that there is a change in the reaction rate constant only when the frequency of a normal-incidence cavity mode is tuned to match a molecular vibrational frequency. Secondly, the effect has to be \emph{collective}, \emph{i.e.} enhanced through the presence of many molecules in the cavity, so that the predicted change in rate constant is appreciable even when the single-molecule coupling is vanishingly small.

So far, only one type of mechanism has been proposed that accurately captures the resonance effect \cite{lindoy2023quantum}; it is rooted in the fact that a correctly-tuned cavity mode can affect the rate of vibrational excitation and de-excitation in the reactive mode, either by coupling to a spectator mode \cite{lindoy2023quantum, ying2025collective,vega2025theoretical}, or directly to the reactive mode itself \cite{lindoy2023quantum, ke2024insights,ke2025stochastic,fiechter2023RPMD,ying2023resonance,ying2024communmater}. Clearly, in the low-friction regime (before Kramers' turnover \cite{peters2017reaction}), this has a direct impact on the overall reaction rate, as in this regime the rate is limited by these very processes. However, these rate changes are only produced in the few-molecule strong-coupling regime; the effect washes out in the limit of many molecules weakly coupled to the cavity mode \cite{lindoy2024investigating,ke2025stochastic}, or else, as soon as one introduces orientational disorder \cite{ying2025collective}.

In this work, we take a different approach. 
We focus on the London dispersion interaction, which is a type of Van der Waals interaction that arises from spontaneous fluctuations in the charge density of neighbouring atoms or molecules \cite{atkinsPC,stone2013theory}. We show how this London dispersion interaction is resonantly modified under VSC, demonstrate the effect this might have on chemical reactivity, and consider how collective effects might come into play. This is especially interesting considering the fact that in several experimental works \cite{joseph2021assembly,patrahau2024direct,imai2025accessing,sandeep2026cluster}, it is hypothesized that altered dispersion interactions are the origin of the reported VSC-induced changes in chemical reactivity.

It has long been known that a change in electromagnetic environment can affect dispersion interactions \cite{buhmann2013dispersion_book}; specifically, results for the interactions between atoms surrounded by two parallel plates have been published as early as in the 1970s \cite{J_Mahanty_1973,minhaeng1996vdW,marcovitch2005enhanced}. 
More recently, dispersion forces have also received attention in the context of polaritonic chemistry. Specifically, studies have shown how intermolecular dispersion interactions are modified under electronic strong coupling (ESC) \cite{haugland2021intermolecular,philbin2023vdWfluids,cao2925cavityinduced,haugland2023understanding}, as well as under VSC \cite{galego2019cavity,fischer2024cavity,sidler2026collectively} in the framework of the cavity Born--Oppenheimer approximation \cite{flick2017cavity,fischer2023beyond,fiechter2024understanding}. 
However, so far none of the effects identified in these studies has been shown to exhibit a sharp resonance behaviour. 

Here, we tackle the problem from a new angle. The basis of our discussion lies in a perturbative treatment of the dispersion interaction. For atoms or molecules that are far enough apart, so that their electron clouds do not overlap, but close enough so that retardation effects can be neglected, the London dispersion interaction can be expressed in second-order perturbation theory as \cite{stone2013theory,buhmann2013dispersion_book}
\begin{equation} \label{eq:eldisp}
     E_\mathrm{LD} = -\sum_{m\neq 0}\sum_{n\neq 0} \frac{\bra{0 0} \hat{V}_\mathrm{d-d}\ket{mn}\bra{mn}  \hat{V}_\mathrm{d-d}\ket{0 0}}{E_{m} - E_{0} + E_{n} -E_{0}},
\end{equation}
where the first index of the ket refers to the electronic state of molecule $A$, and the second to that of molecule $B$.
The intermediate states are electronic excited states, and the perturbation is the dipole--dipole interaction, which (in atomic units) is given by
\begin{equation}
\begin{aligned}
    \hat{V}_\mathrm{d-d} &= \frac{\hat{\boldsymbol{\mu}}_A \cdot \hat{\boldsymbol{\mu}}_B-3(\hat{\boldsymbol{\mu}}_A\cdot \hat{r})(\hat{\boldsymbol{\mu}}_B\cdot \hat{r})}{r^3} \\
\end{aligned}
\end{equation}
Here, $\hat{\boldsymbol{\mu}}_A$ and $\hat{\boldsymbol{\mu}}_B$ are the dipole moment operators for molecule $A$ and $B$, respectively, $r$ is the distance between the molecules, and $\hat{r}$ the unit vector along the intermolecular axis. As $\hat{V}_\mathrm{d-d} \propto 1/r^3$ is featured twice in Eq.~\eqref{eq:eldisp}, clearly $E_\mathrm{LD}\propto 1/r^6$, as is characteristic of Van der Waals interactions.

From Eq.~\eqref{eq:eldisp} it is not obvious how VSC might resonantly affect dispersion forces -- as the expression features no vibrations, it is unclear how hybridization of these vibrations with a cavity mode would have any effect on the dispersion interaction. One could try replacing the electronic excited states in Eq.~\eqref{eq:eldisp} with vibrational excited states, to obtain ``vibrational dispersion forces". However, these are orders of magnitude smaller than the standard electronic London dispersion forces. As we will see in the next section, the key here lies instead in resolving the vibrational states within the electronic levels. 

\section{Results}
\subsection{Vibrationally-resolved London dispersion interactions}

\begin{figure}
    \centering
    \includegraphics[width=0.8\linewidth]{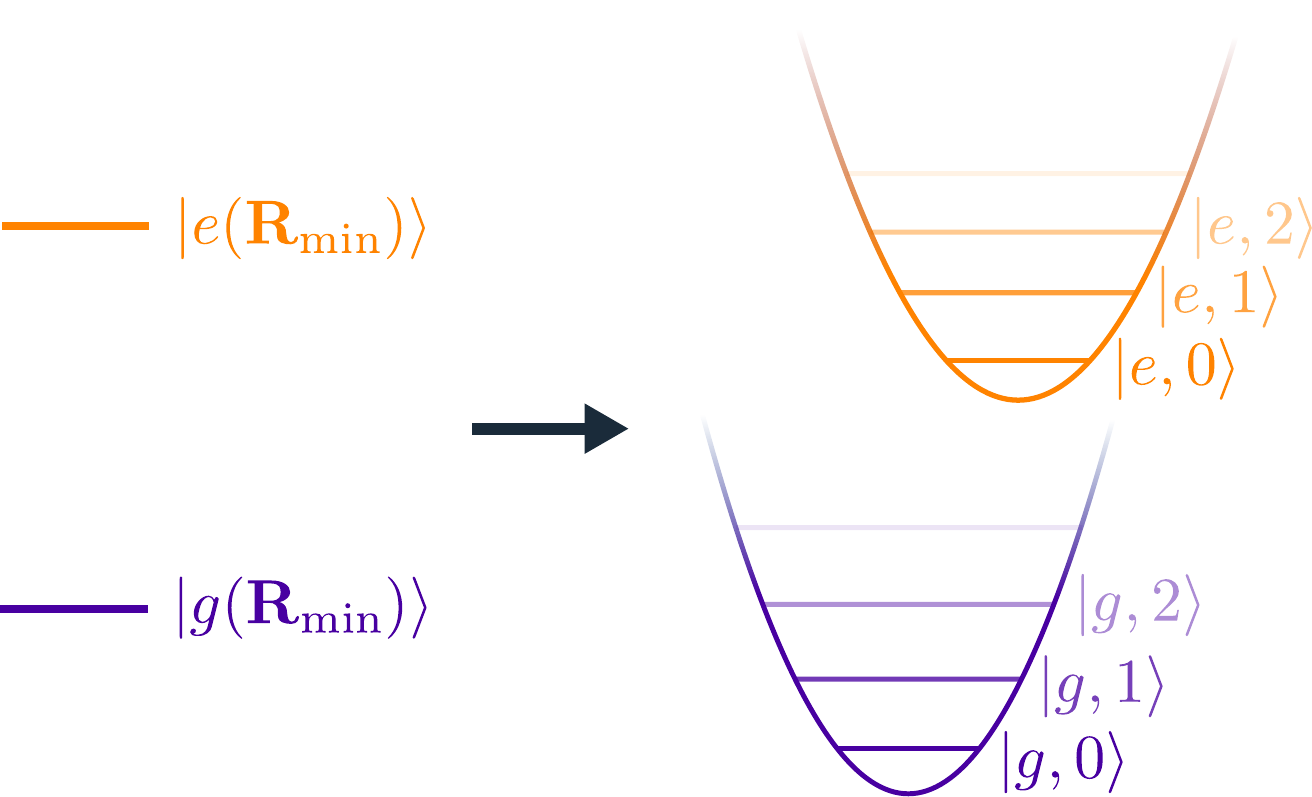}
    \caption{\textbf{Vibrationally-resolved London dispersion forces.} Instead of calculating the dispersion interaction based on electronic states at a fixed nuclear configuration (Eq.~\eqref{eq:eldisp}), we will calculate it for a given vibrational state (Eq.~\eqref{eq:vibdisp}). }
    \label{fig:vibres}
\end{figure}

In order to capture the effect of \emph{vibrational} strong coupling on the London dispersion interaction, we incorporate vibrational states in our description, as illustrated in Fig.~\ref{fig:vibres}. For simplicity, from here on we only consider two electronic states per molecule, $e$ and $g$, separated by an energy $\hbar\omega_{eg}$. Additionally, we will consider one vibrational degree of freedom per molecule for now. Quantizing this vibration leads to the following dispersion energy, which now depends on the vibrational states $\alpha$ and $\beta$ for molecules $A$ and $B$, respectively,
\begin{equation} \label{eq:vibdisp}
E_\mathrm{LD}(\alpha, \beta) = -\sum_{\rho,\sigma} \frac{\big|\bra{g,\alpha; g,\beta} \hat{V}_\mathrm{d-d} \ket{e,\rho;e,\sigma}\big|^2 }{2\hbar \omega_{eg} + E_\rho - E_\alpha + E_\sigma - E_\beta}. 
\end{equation}
Note that the sum over intermediate states now runs over all vibrational states $\rho$ and $\sigma$ in the electronic excited state of molecules $A$ and $B$, respectively. Also, as typically the electronic excitation energy $\hbar\omega_{eg}$ is much larger than vibrational energies (here $E_\rho$, $E_\sigma$, $E_\alpha$, $E_\beta$), the denominator can simply be approximated by $2\hbar\omega_{eg}$; we will drop the slight dependence of the denominator on the vibrational states for the remainder of this work. 

For the sake of brevity, we will simplify Eq.~\eqref{eq:vibdisp} further by assuming that molecules $A$ and $B$ are oriented such that their electronic transition dipole moments are aligned, and perpendicular to the intermolecular axis (note that it is in principle straightforward to bring the orientation-dependence of the dispersion interaction back into the description later). This results in the compact expression, 
\begin{equation}\label{eq:C6nocav}
\begin{aligned} 
    E_\mathrm{LD}(\alpha,\beta) \approx& - \frac{1}{r^6} \frac{1}{2\hbar \omega_{eg}} \sum_{\rho,\sigma} \big|\bra{g,\alpha; g,\beta} \hat{\mu}_A \hat{\mu}_B \ket{e,\rho;e,\sigma}\big|^2 \\
    = & - \frac{1}{r^6} \frac{1}{2\hbar \omega_{eg}} \bra{\alpha; \beta} \mu_{eg}(\hat{x}_A)^2\mu_{eg}(\hat{x}_B)^2 \ket{\alpha; \beta}\\
    \equiv& - \frac{1}{r^6} \frac{1}{2\hbar\omega_{eg}} \mathrm{M}_{\alpha \alpha} \mathrm{M}_{\beta \beta}\\
    \equiv& -\frac{C_6(\ket{\alpha;\beta})}{r^6}
    \end{aligned}
\end{equation}
upon which we will build. Going from the first to the second line, we have used the fact that the vibrational states form a complete set, so that $\sum_\rho \ket{\rho}\bra{\rho}=\mathbb{1}$; the same goes for the states labelled with $\sigma$. We have also introduced the nuclear position operators $\hat{x}_A$ and $\hat{x}_B$ of molecules $A$ and $B$, and the abbreviated notation
\begin{equation} \label{eq:matrixelement}
    \mathrm{M}_{\alpha\alpha} = \bra{\alpha} \mu_{eg}(\hat{x})^2 \ket{\alpha}
\end{equation}
where $\hat{\mu}_{eg} = \hat{\mu}_{ge} = \bra{e}\hat{\mu}\ket{g}$ is the electronic transition dipole moment, which depends on the nuclear position. For a given functional form of $\mu_{eg}(x)$ and of the wavefunction $\psi_\alpha(x)$ of vibrational state $\alpha$, one can calculate these matrix elements by simply doing the integral over $x$. For the setup used in the majority of calculations reported in this study, we find $C_6(\ket{0,0})=\mathrm{M}_{00}^2/2\hbar\omega_{eg}=1.65$ a.u. and $C_6(\ket{1,0})=\mathrm{M}_{00}\mathrm{M}_{11}/2\hbar\omega_{eg}=2.35$ a.u. (setting $2\hbar\omega_{eg}=1$ a.u. for convenience);  
this means that the London dispersion interaction between $A$ and $B$ increases slightly upon vibrationally exciting one of the molecules. More details can be found in the Methods section.

\subsection{London dispersion interaction under vibrational strong coupling} \label{sec:LD_VSC}

\begin{figure}
    \centering
    \includegraphics[width=0.65\linewidth]{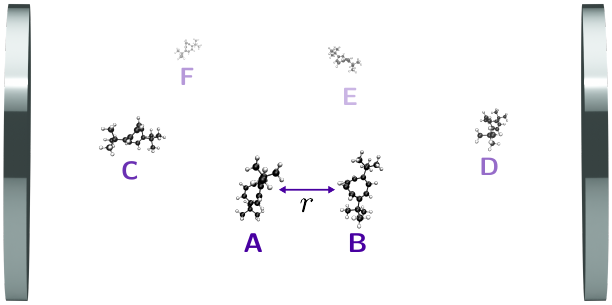}
    \caption{\textbf{Setup of the system.} Molecules $A$ and $B$ are in close vicinity of each other, and interact with each other via dispersion; we assume their electronic transition dipole moments to be aligned and perpendicular to the intermolecular axis, and their ground-state dipole moments to be aligned and parallel to the mirrors, for efficient coupling to a normal-incidence cavity mode. The other molecules can be anywhere in the cavity; they interact with $A$ and $B$ only via the cavity mode. }
    \label{fig:moleculesincav}
\end{figure}

\begin{figure*}
    \centering
    \includegraphics[width=\linewidth]{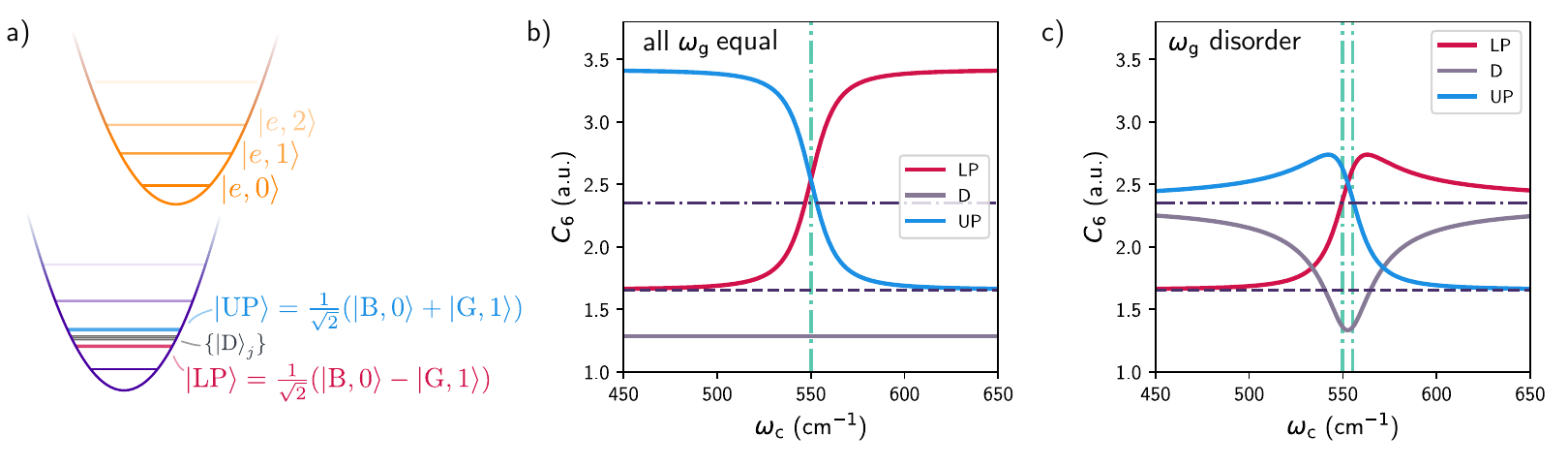}
    \caption{\textbf{Vibrational strong coupling modifies the London dispersion interaction.} a) States relevant for the calculation of the dispersion interaction. When the cavity is tuned to the vibrational frequency (of the electronic ground state), polaritons and dark states form. b) The state-dependent $C_6$ coefficient (setting $2\hbar\omega_{eg}=1$ a.u. for convenience) for the dispersion interaction between molecules $A$ and $B$, plotted as a function of cavity frequency for the case of $N=2$ molecules, both with $\omega_g=$ \SI{550}{\per\centi\meter} (light green vertical line), and at a coupling strength of $g_\mathrm{c}=$ \SI{5}{\per\centi\meter}. 
    On resonance, in both the upper and lower polaritonic states (blue and red line, respectively) the dispersive forces are stronger than outside the cavity (\emph{c.f.} the horizontal black dashed and dashed-dotted lines: the $C_6$ values for both $A$ and $B$ in the vibrational ground state, and for one of them vibrationally excited, respectively). 
    In contrast, when in the dark state, molecules $A$ and $B$ experience weaker dispersion interactions. 
    c) Similar to b), but now each molecule has a slightly different $\omega_g$ (light green vertical lines). Importantly, in contrast to b), all $C_6$ coefficients tend to an out-of-cavity value when the cavity is tuned away from resonance. Nevertheless, on resonance, it is still enhanced compared to the out-of-cavity value. }
    \label{fig:VSC}
\end{figure*}

We are now in a position to investigate how a cavity mode can affect these vibrationally-resolved London dispersion interactions. We consider a slightly more general situation, illustrated in Fig.~\ref{fig:moleculesincav}: we place $N$ molecules in the cavity. Only molecules $A$ and $B$ are in close vicinity of each other, and they interact via dispersion forces; the other molecules can be far away, and only interact with $A$ and $B$ via the cavity mode. For simplicity, we will consider strong coupling only to vibrations in the electronic ground state (and not to the vibrations in the electronic excited state). Also, in this section, we assume all molecules to be coupled to the cavity mode with equal coupling strength. Later, we show that going beyond this by accounting for the spatial variation of the mode profile and orientational disorder does not qualitatively affect our final conclusions.

Tuning the cavity frequency $\omega_\mathrm{c}$ close to a vibrational frequency $\omega_g$ of the molecules leads to the formation of new eigenstates $\ket{j}$ (Fig.~\ref{fig:VSC}a). This in turn affects the London dispersion interaction; the modified $C_6$-coefficient is given by 
\begin{equation} \label{eq:C6general}
C_6(\ket{j} ) =  \frac{1}{2\hbar \omega_{eg}} \bra{j} \mu_{eg}(\hat{x}_A)^2\mu_{eg}(\hat{x}_B)^2 \ket{j}.
\end{equation}
Via a basis transformation, it is straightforward to relate these matrix elements to the matrix elements in the original basis (\emph{c.f.} Eq.~\eqref{eq:matrixelement}, see also Methods). 

For now, it is instructive to assume all molecules to have the same vibrational frequency; this allows us to use analytical expressions for the eigenstates in our investigations. Following the Tavis--Cummings model \cite{tavis1968exact} (see Methods), the first excitation manifold contains two polaritonic states, the lower and upper polariton,
\begin{subequations} \label{eq:LPUP}
\begin{align}
    \ket{\mathrm{LP}} &= -\sin\Theta \ket{B} \otimes \ket{0}_\mathrm{c} + \cos \Theta \ket{G} \otimes \ket{1}_\mathrm{c} \\
    \ket{\mathrm{UP}} &= \cos\Theta \ket{\mathrm{B}} \otimes \ket{0}_\mathrm{c} + \sin \Theta \ket{\mathrm{G}} \otimes \ket{1}_\mathrm{c},
\end{align}
\end{subequations}
respectively.
These states are mixed light--matter excitations, composed in part of an excitation in the cavity mode $ \ket{1}_\mathrm{c}$ combined with all molecules in their vibrational ground state $\ket{\mathrm{G}}$, and in part of the cavity mode in its ground state $\ket{0}_\mathrm{c}$, combined with the ``bright state" $\ket{\mathrm{B}}$. This bright state is a symmetric superposition of all states $\ket{1}_i$, in which a single molecule $i$ is in its vibrational excited state, while the others remain in their ground state:
$   \ket{\mathrm{B}} = \frac{1}{\sqrt{N}} \sum_{i=1}^N \ket{1}_i $. 
The extent of the mixing depends on the cavity detuning $\omega_\mathrm{c}-\omega_g$, and is set by the mixing angle $\Theta=\tfrac{1}{2}\arctan ( 2 g_\mathrm{c} \sqrt{N}/(\omega_\mathrm{c}-\omega_g) )$, where $g_\mathrm{c}$ is the light--matter coupling strength; on resonance, $\sin\Theta=\cos\Theta=1/\sqrt{2}$. The other $N-1$ states in the single-excitation subspace are ``dark states" $\ket{\mathrm{D}}_j$, \emph{i.e.} antisymmetric linear combinations of molecular excitations \cite{mandal2023theoretical}.

We start by investigating the case of $N=2$ molecules in the cavity, with results summarized in Fig.~\ref{fig:VSC}. In this case, the antisymmetry of the dark state weakens the dispersive forces,
\begin{equation}
C_6(\ket{\mathrm{D}})=\frac{1}{2\hbar \omega_{eg}}\big( \mathrm{M}_{11} \mathrm{M}_{00} - \mathrm{M}_{01}^2 \big),
\end{equation}
while the symmetry of the bright state leads to enhanced dispersion interactions, 
\begin{equation}
C_6(\ket{\mathrm{B}})=\frac{1}{2\hbar \omega_{eg}}\big( \mathrm{M}_{11} \mathrm{M}_{00} + \mathrm{M}_{01}^2 \big).
\end{equation}
This is partially inherited by the polaritonic states,
\begin{equation} \label{eq:C6polariton}
C_6(\ket{\mathrm{LP}}\text{ or }\ket{\mathrm{UP}})=\frac{1}{4\hbar \omega_{eg}}\big( \mathrm{M}_{11} \mathrm{M}_{00} + \mathrm{M}_{01}^2  + \mathrm{M}_{00}^2\big).
\end{equation}
These observations are reflected in Fig.~\ref{fig:VSC}b, where we  plot $C_6$ as a function of frequency. Indeed, $C_6$ is suppressed for the dark state, to a value even lower that of the ground state. In contrast, both polaritonic states experience an enhanced $C_6$ when the cavity is tuned to resonance, due to their bright-state character. Strikingly, when moving away from resonance, $C_6$ is enhanced even further for one of the polaritons, following the fact that this state gains more bright-state character as the detuning increases.

With an eye on the next section, it is both more realistic and more convenient to incorporate frequency disorder (\emph{i.e.} all molecules having a slightly different $\omega_g$, due to each of them having a slightly different local environment) in the description (see Methods). The effect of this on $C_6$ is shown in Fig.~\ref{fig:VSC}c for a particular realization of disorder. Close to resonance, we still clearly see an enhancement of $C_6$ for the polaritonic states. An important consequence of frequency disorder emerges at large detuning: since the single-molecule excited states are no longer degenerate, they do not form bright and dark states far off-resonance. Instead, all states tend to localized singly-excited states, and their $C_6$ coefficients reduce to their out-of-cavity value accordingly.

\subsection{Resonant acceleration of a simple reaction} \label{sec:reactionN2}

\begin{figure*}
    \centering
    \includegraphics[width=\linewidth]{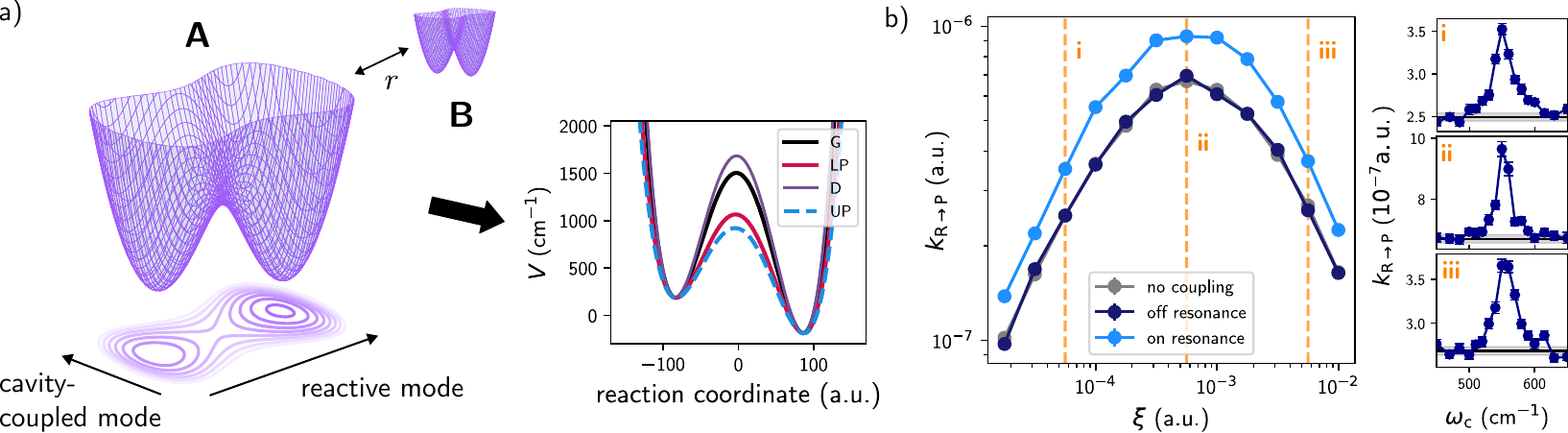}
    \caption{\textbf{A simple model of a reaction influenced by dispersion forces.} a) Outline of the model: we consider the case of $N=2$ molecules, with each molecule consisting of two modes, one of which is coupled to the cavity, the other being a reactive mode represented by a double-well potential. We will focus on the reaction in molecule $A$. We integrate out the cavity mode (set to $\omega_\mathrm{c}=$ \SI{550}{\per\centi\meter}) and cavity-coupled molecular modes, which leaves us with a series of potential energy surfaces along $A$'s reaction coordinate. The barrier height of each of these surfaces depends on the strength of the dispersion interaction between $A$ and $B$, and thereby on which state the system is in. b) Kramers' plot of the forward reaction rate constant $k_\mathrm{R\rightarrow P}$ vs. the Langevin friction $\xi$, showing the typical turnover behaviour. Coupling to a cavity mode ($g_\mathrm{c}=$ \SI{5}{\per\centi\meter}, other parameters listed SI) leads to resonant rate enhancement across all friction regimes, from low friction (i), to turnover (ii), to high friction (iii) (\emph{c.f.} the gray shaded area, which is the out-of-cavity rate). The error bars represent the 68\% confidence interval; in the Kramers' plot, the marker size exceeds the error bar size. }  
    \label{fig:model}
\end{figure*}

In the previous section, we have demonstrated that the state-dependent dispersion interaction between two molecules is modified when tuning a cavity to resonance with a vibration. Nevertheless, in the end we are interested in studying a reaction mixture that is at thermal equilibrium, so that the population is distributed over the vibrational states, rather than all being in one specific excited vibrational state. 
A good starting point might therefore be the thermally averaged value of the $C_6$ coefficient, $\expval{C_6} = \tfrac{1}{Z}\sum_i \mathrm{e}^{-\beta E_i} C_{6,i}$, where $Z$ is the partition function, and the sum runs over vibrational/vibropolaritonic eigenstates $i$ with energy $E_i$. However, as shown in the SI, coupling a cavity does not affect this thermally averaged $\expval{C_6}$. Note that this thermal-averaging procedure completely disregards dynamical effects: it effectively assumes that thermalization is instantaneous, \emph{i.e.} that hopping between ground, polaritonic and dark states is much faster than any other time scale in the system. This is not true for parameters typical of experiment. As we show below, dynamical effects turn out to play a key role in the resonant modification of chemical dynamics.

In order to study how cavity-modified London dispersion forces can translate to changes in reactivity, we develop a simple model of a chemical reaction (Fig.~\ref{fig:model}). Details and parameters can be found in the Methods section; here we give a summary. In our setting, each molecule has two vibrational modes: a reactive mode, characterized by an asymmetric double-well potential (representing \emph{e.g.} an isomerization reaction or rearrangement); and a vibrational mode perpendicular to it, to which the cavity mode is coupled (Fig.~\ref{fig:model}a). We focus on the reactivity of molecule $A$, and assume that the barrier height in its reactive mode is influenced by the interaction with the nearby molecule $B$. In our model, stronger London dispersion interactions between $A$ and $B$ stabilize $A$'s transition state and lead to a lowering of the barrier. 

We proceed by quantizing the perpendicular vibrational modes and cavity mode (Fig.~\ref{fig:model}b). This leads to a state-dependent potential along the reaction coordinate: when the system as a whole is in a polaritonic state, $A$'s barrier height is lowered, while in the dark state $A$ has to overcome a higher barrier to react. 

To evaluate the reaction rate constant, we run classical Langevin dynamics along the reaction coordinate, while hopping between states according to a Pauli master equation \cite{breuer2002theory,may2023charge,du2022catalysis}. All trajectories are started in a quasi-equilibrium distribution on the left of the barrier (defined as the `reactant' region), and are counted as products after passing the top of the ground-state barrier. After some time, the dynamics of the product population settle down into rate-like behaviour, and relax exponentially towards its equilibrium value. The forwards and backwards rate constants are extracted from an exponential fit to the data (excluding the initial transient behaviour); this is illustrated with an example in the SI. The error bars show the 68\%  confidence interval obtained from a bootstrapping procedure. 

In Fig.~\ref{fig:model}b, we show the rate constant as a function of Langevin friction for an on-resonance cavity ($\omega_\mathrm{c}=$ \SI{550}{\per\centi\meter}), off-resonance cavity ($\omega_\mathrm{c}=$ \SI{450}{\per\centi\meter}), and without coupling to a cavity. Clearly, across the whole range of frictions, coupling to a cavity mode resonantly enhances the reaction rate in this model. Note that we use the frequency disorder introduced in the previous section, which is of major importance here: without it, we would be breaking the secular approximation that underlies the Pauli master equation we are using. 

This rate enhancement arises in part from the barrier lowering: hopping to a polaritonic state lowers the barrier more than hopping to a bare excited vibrational state would. However, there is an additional factor that plays a role: the lifetime difference between polaritons and dark states/bare vibrations. On resonance, the polaritonic lifetime is given by the arithmetic average of the cavity lifetime $\kappa$ and the vibrational lifetime $\gamma$. As typically, $\kappa \gg \gamma$, the lifetime of polaritonic states is much shorter than that of out-of-cavity vibrational states; by detailed balance, this also means that they are visited more often. Trajectories can therefore more frequently profit from barrier lowering that enhances the rate. Indeed, as shown in the SI, increasing the cavity's $Q$-factor, and thereby increasing its lifetime, leads to a decrease in the rate enhancement.





\subsection{Collective effects}


So far, we have focused on the case of $N=2$ molecules. However, in experiment, typically millions of molecules are simultaneously coupled to the same cavity mode. While it is infeasible to obtain numerical results in this many-molecule collective limit, it is instructive to try and extract a trend in the few-molecule regime.

\begin{figure}
    \centering
    \includegraphics[width=\linewidth]{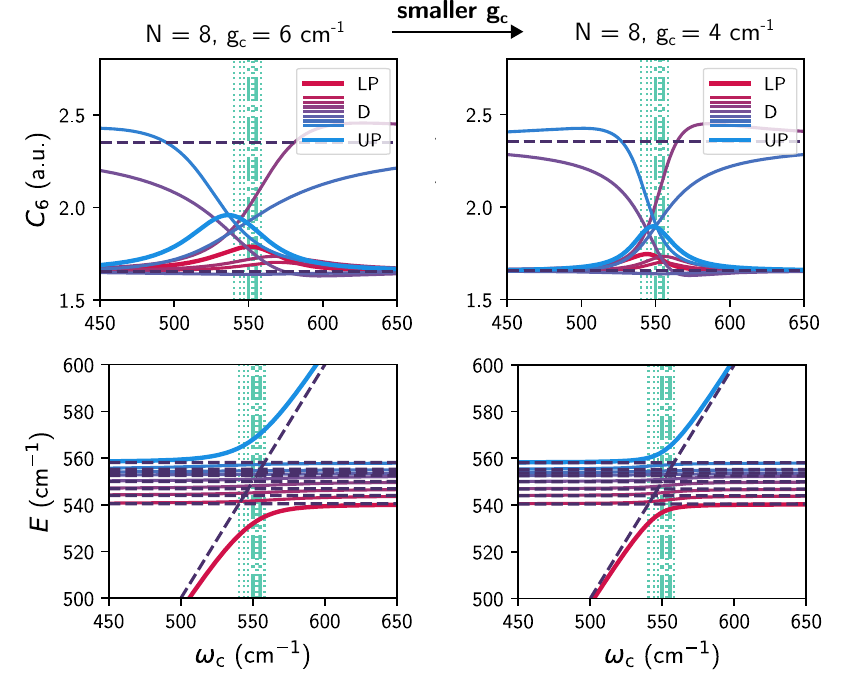}
    \caption{\textbf{Cavity-frequency dependence of the $C_6$ coefficient and the eigenstate energies with 8 molecules in the cavity}, for two different coupling strengths. Molecules $A$ and $B$ have the same frequency as in Fig.~\ref{fig:VSC}c (green dash-dotted vertical lines), the frequencies of the other molecules are indicated with green dotted vertical lines. The thicker blue and red lines are the upper and lower polaritonic state, respectively. Unlike in the two-molecule case, their $C_6$ is never enhanced above out-of-cavity levels (black dashed lines).  Decreasing the coupling strength $g_\mathrm{c}$ contracts the width of the features in the plot. Close to resonance however, the extent to which the cavity changes the $C_6$ coefficient is similar irrespective of coupling strength.   }
    \label{fig:C6_8mol}
\end{figure}

Let us start out by considering the behaviour of the dispersion interaction between molecules $A$ and $B$ when the entire system is in one of the polaritonic states. For the case without disorder, where analytical expressions for the wavefunctions are available (Eq.~\eqref{eq:LPUP}), and with the cavity tuned to resonance, the $C_6$ coefficient that characterizes this dispersion interaction is given by
\begin{equation}
\begin{aligned}
C_6(\ket{\mathrm{LP}}\text{ or }\ket{\mathrm{UP}})=\frac{1}{2\hbar \omega_{eg}}\Big(\frac{1}{N} \big[\mathrm{M}_{11} \mathrm{M}_{00} + \mathrm{M}_{01}^2 \big] \\
+ \mathrm{M}_{00}^2\Big[1-\frac{1}{N}\Big]  \Big).
\end{aligned}
\end{equation}
This means that the polaritonic states behave more and more like the ground state (for which $C_6=(2\hbar\omega_{eg})^{-1} \mathrm{M}_{00}^2$) as $N\rightarrow \infty$, washing away the interesting features we saw in the $N=2$ case. Physically, the reason for this is that the excitation is spread out all over the ensemble; adding more molecules effectively dilutes the excitation on molecules $A$ and $B$. This trend can also be observed for the case with frequency disorder. For example, from Fig.~\ref{fig:C6_8mol}, where the $C_6$ coefficients are plotted as a function of $\omega_\mathrm{c}$ for $N=8$ molecules, it is clear that dispersion interactions of the polaritonic states are weaker than those of the out-of-cavity vibrationally excited states. This stands in contrast to the situation for $N=2$ molecules plotted in Fig.~\ref{fig:VSC}c, where the polaritonic $C_6$ coefficients exceed the out-of-cavity value when the cavity is tuned to resonance. 

Another difference with the $N=2$ situation is that away from resonance, the polaritonic $C_6$ coefficients do not tend to the out-of-cavity excited state values. The underlying reason for this is that now, the molecules $A$ and $B$ do not have the lowest and highest frequency in the ensemble anymore. Because of this, when moving off-resonance to higher frequencies, the lower-polariton excitation localizes on the lowest-frequency oscillator; conversely, when moving to lower frequencies the upper-polariton excitation localizes on the highest-frequency oscillator -- this is also clear (albeit indirectly) from the lower two panels of Fig.~\ref{fig:C6_8mol}. If, as in our case, neither of these oscillators are molecule $A$ or $B$, then the polaritonic $C_6$ tends to that of the ground state when moving off-resonance.

In Fig.~\ref{fig:C6_8mol} we show more generally how the $C_6$ vs.~$\omega_\mathrm{c}$ profile changes as we adjust the light--matter coupling strength $g_\mathrm{c}$. Decreasing the coupling strength leads to a contraction in the features; \emph{i.e.}~when detuning the cavity away from resonance, the $C_6$ coefficients return more quickly to their out-of-cavity values. Interestingly, a lower coupling strength does not necessarily translate to less cavity-induced changes around resonance -- in both $C_6$ plots in Fig.~\ref{fig:C6_8mol}, the $C_6$ coefficients on resonance are very similar. Decreasing the coupling strength merely narrows the range of frequencies that can be considered to be ``on resonance". The underlying reason for this behaviour is that the $C_6$ coefficient is a property that depends directly on the polaritonic or dark state wavefunction (\emph{c.f.} Eq.~\eqref{eq:C6general}), instead of \emph{e.g.} on the energy of the polaritonic states (which via the Rabi splitting scales with $g_\mathrm{c}$; see the lower panels in Fig.~\ref{fig:C6_8mol}).

\begin{figure}
    \centering
    \includegraphics[width=0.7\linewidth]{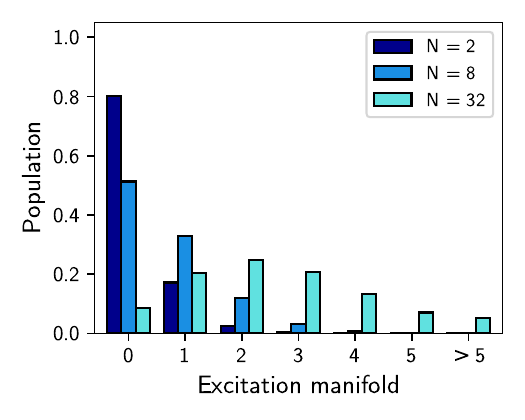}
    \caption{\textbf{Higher excitation manifolds become populated when increasing the number of molecules $N$}, here for simplicity shown for $g_\mathrm{c}= $ \SI{0}{\per\centi\meter} in the absence of frequency disorder, meaning $\omega = $ \SI{550}{\per\centi\meter} for all oscillators; we set $T=$ \SI{300}{\kelvin}. The population of the $n$th excitation manifold is given by $g_n \mathrm{e}^{-\beta E_n}/Z$, where $E_n=n\hbar\omega$, $Z$ is the total partition function, and $g_n$ the number of states with $n$ excitations. It is this degeneracy $g_n$ that grows quickly with $N$.}
    \label{fig:exman_population}
\end{figure}

\begin{figure*}
    \centering
    \includegraphics[width=0.9\linewidth]{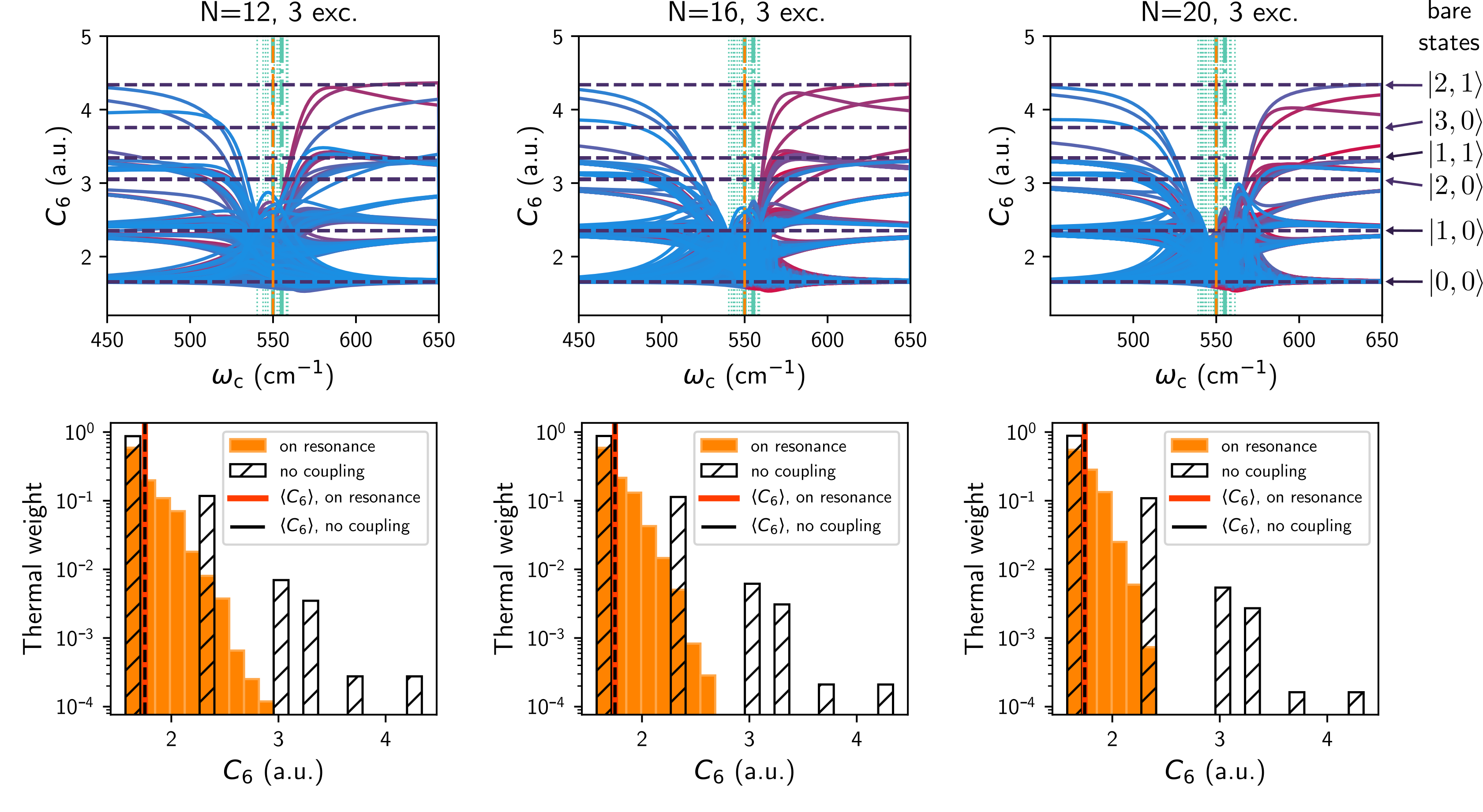}
    \caption{\textbf{Behaviour of the state-dependent $C_6$ coefficients for increasing numbers of molecules.} Top row: cavity-frequency-dependence of the $C_6$ coefficients for 12, 16 and 20 molecules coupled to the cavity mode, respectively. Only states with a total of 3 excitations or fewer are included. The dashed black lines indicate the relevant out-of-cavity $C_6$ coefficients, and on the right we show the states they correspond to, in terms of the number of vibrational excitations $n_A$ and $n_B$ of molecules $A$ and $B$: $\ket{n_A,n_B}$ (note that $C_6(\ket{n_A,n_B})=C_6(\ket{n_B,n_A})$; only one state is shown for the sake of brevity). $g_\mathrm{c}$ is set to \SI{5}{\per\centi\meter} for $N=12$, and multiplied by $\sqrt{12/N}$ when increasing the number of molecules, to keep the Rabi splitting approximately constant. As before, all molecules have slightly different frequencies (green vertical lines; dash-dotted for $A$ and $B$, dotted for the other molecules). On top of that, we also accounted for orientational and positional disorder of all molecules except $A$ and $B$ by adjusting their $g_\mathrm{c}$ (see Methods). A key observation is that on resonance (at \SI{550}{\per\centi\meter}, orange vertical line in top row figures), there is a ``dip" in the plot: there are no states with a $C_6$ coefficient as high as outside the cavity. This feature persists when increasing the number of molecules, making it a collective effect.
    Bottom row: the distribution of $C_6$ coefficients at \SI{550}{\per\centi\meter}, weighted by the thermal population; this quantifies the ``dip" in $C_6$ coefficients around resonance evident from the top-row figures. As in the 2-molecule case, the mean of the distribution, $\expval{C_6}$, is not affected by the cavity (red / black dashed lines). However, the cavity does affect the width of the distribution; for all numbers of molecules it is consistently narrower than outside the cavity.    }
    \label{fig:C6_manymol}
\end{figure*}

However, this is not the whole story. So far, we have only been considering the ground state and the single excitation manifold. As we increase $N$ though, it becomes more and more likely that the molecules have more than one excitation shared amongst them. As shown in Fig.~\ref{fig:exman_population}, indeed higher excitation manifolds become populated for higher $N$. The shape of the population diagram is reminiscent of a the population distribution over rotational states, as both originate from the competition between energetics and degeneracy: for larger $N$, the degeneracy of the higher excitation manifolds grows quickly (see Methods for the expression), effectively pulling population up from the ground state and lower manifolds.

Note that already for $N=8$, it seems that we should allow for at least two, perhaps three excitations in the whole system -- the maximum number of excitations thus becomes a new convergence parameter. At this point, it is relevant to note that we will treat each molecule as a harmonic oscillator, allowing more than one excitation per molecule (as opposed to the Tavis--Cummings model, which assumes atoms or molecules to be two-level systems). Not only is this closer to reality, it also provides some insight in the eigenstates: for a system of coupled harmonic oscillators, these can be expressed in terms of the number of excitations per eigenmode, which in this case are the polaritonic and the dark modes \cite{du2022catalysis}. This structure is also useful in general when computing decay rates for each state, as shown in the SI.

With this in mind, we move on to studying the cavity-modification of the $C_6$ coefficient for increasing numbers of molecules in the cavity. The aim here is to extract a trend that we can extrapolate to the many-molecule collective limit.
To make the setting more realistic, and in contrast to the results for $N=8$ in Fig.~\ref{fig:C6_8mol}, we now also account for variations in $g_\mathrm{c}$. This disorder in $g_\mathrm{c}$ arises from each molecule having a different orientation compared to the cavity field's polarization, and from the molecules feeling a different electric field strength at different positions in the cavity; details can be found in the Methods section. We keep the position of molecules $A$ and $B$ fixed at an antinode of the cavity mode, and fix their orientations too for simplicity -- especially as changing their relative orientation would not only change the extent to which they couple to the cavity, but also affect their dispersion interaction.

In Fig.~\ref{fig:C6_manymol}, we present results for 12, 16 and 20 molecules in the cavity, including all states that have 3 excitations or fewer. In the $C_6$-vs.-$\omega_\mathrm{c}$ plots, the main feature that stands out is the ``dip" around resonance: here, there are no states available in which the dispersion interaction between $A$ and $B$ is as large as it can be outside the cavity. This is shown more clearly in the figures in the bottom row, where we plotted the distribution of the states' $C_6$ coefficients, weighted by their thermal populations. As in the 2-molecule case, coupling to a cavity does not alter $\expval{C_6}$, the mean of the distribution. However, in a resonant cavity, the distribution is much more compact than in the out-of-cavity case: while there are many states with a $C_6$ that is elevated slightly above that of the vibrational ground state, the states with higher $C_6$ that are populated outside the cavity are absent in the `on resonance' case. 

\begin{figure}
    \centering
    \includegraphics[width=0.75\linewidth]{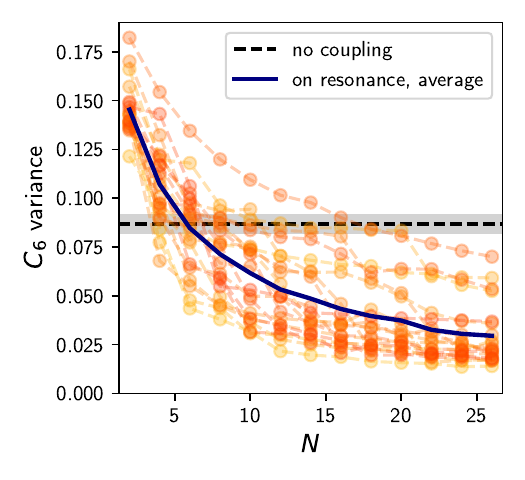}
    \caption{\textbf{The variance of the thermal distribution of $C_6$ coefficients for increasing numbers of molecules.} Each shade corresponds to a different configuration of the system, with randomly selected molecular frequencies, orientations, and positions in the cavity. $g_\mathrm{c}=$ \SI{20}{\per\centi\meter} for $N=2$, and multiplied by $\sqrt{2/N}$ as we increase $N$. The out-of-cavity variance also varies slightly with disorder configuration, as the frequencies of molecules $A$ and $B$ affect the wavefunctions that go into Eq.~\eqref{eq:C6nocav}. We found that allowing for up to 4 excitations in the system leads to converged results -- at least for $N<16$; beyond this including states with 5 excitations becomes computationally infeasible. The average on-resonance variance converges to a value significantly below the no-coupling variance, suggesting that this effect may persist in the collective limit. }
    \label{fig:Ndependence}
\end{figure}

To quantify this narrowing of the distribution, we calculate its variance, $\mathrm{var} \,C_6 =\tfrac{1}{Z}\sum_i \mathrm{e}^{-\beta E_i} (C_{6,i} - \expval{C_6})^2 $. In Fig.~\ref{fig:Ndependence}, we show the results for 20 different configurations of the system, \emph{i.e.} with randomly selected vibrational frequencies, molecular orientations, and positions in the cavity.  For low numbers of molecules, and when coupled to an on-resonance cavity, the variance of the thermally-weighted $C_6$ coefficient exceeds the out-of-cavity value -- this is in qualitative agreement with our observations for the $N=2$ case in Fig.~\ref{fig:VSC}c. When increasing the number of molecules coupled to the cavity mode, the variance drops significantly, and for every disorder realization, it ends up below the out-of-cavity variance by $N=26$, after which point our calculations become prohibitively expensive. However, the trend seems clear, with the effect flattening out and persisting with increasing $N$. This suggests it is a collective effect that has good chances to survive in the large-$N$ limit ($N\sim 10^6$ or more) relevant to experiment.

Some intuition behind the origin of this effect might be found in the ``semilocalization" property of dark states: for large enough $N$ and when including frequency disorder, an excitation of a dark mode is typically delocalized across a few ($\sim 2-3$) molecules \cite{dark2020botzung,du2022catalysis}. 
Exciting such a dark mode therefore leads to a smaller increase in $C_6$ coefficient than exciting the localized vibration of molecule $A$ or $B$ would. To reach a $C_6$ coefficient similar to that of a bare excited state, one would have to excite multiple specific dark modes, which is statistically less likely to occur.

So far, we have discussed the collective modification of the dispersion interaction. Nevertheless, in the end we are interested in whether a cavity can collectively modify reaction rates. This is clearly a possibility: in an on-resonance cavity, fewer states are available in which molecules $A$ and $B$ experience a strongly enhanced dispersion interaction. If outside the cavity, the reaction strongly depends on molecule $B$ stabilizing molecule $A$'s transition state via these enlarged dispersion forces in vibrationally excited states (cf. Fig.~\ref{fig:model}a), then coupling a cavity mode could suppress the rate. However, it is unfortunately not possible for us to test this hypothesis in the mixed quantum--classical framework we have been using. The reason for this is that the Pauli master equation that underlies our approach hinges on the secular approximation, which assumes that all quantum states are energetically well-separated (compared to their linewidths \cite{du2022catalysis}). When restricting the calculation to the first excitation manifold, this condition is easily satisfied by choosing the molecular frequencies to be well-separated, as states do not cross (as a function of $\omega_\mathrm{c}$, illustrated in the SI). However, as discussed above, increasing the number of molecules means that we also need to include higher excitation manifolds, in which states do cross. Additionally, there are more states that have some cavity-excitation character mixed in, leading to line broadening. Altogether, this leads to the breaking of the secular approximation around resonance, thereby invalidating our mixed quantum--classical dynamics scheme there. Exploring rate changes for a large number of molecules is thus not possible within our current framework; we leave the development of a new approach that can capture this regime for future work.

\section{Discussion}

In this work, we have shown how VSC can resonantly modify vibrationally-resolved London dispersion forces. We have also presented evidence suggesting that such a modification persists in the collective limit and in the presence of orientational disorder. This is relevant to the experiments that use a planar microcavity, in which a large number of molecules are each weakly coupled to a cavity mode. 

Based on a mixed quantum--classical dynamics scheme, we have demonstrated that in the case of two molecules strongly coupled to a cavity mode, these VSC-modified dispersion interactions can give rise to resonant enhancement of the reaction rate constant. In contrast to earlier theoretical studies that found resonant acceleration of the reaction only in the low friction regime \cite{lindoy2023quantum}, our results show sharp peaks also in the high-solvent-friction regime, which is likely to be more relevant for solution-phase chemistry \cite{nitzan2006chemical}.




As this is a new line of inquiry, many open questions still need to be addressed. 
For example, what parameter ranges are realistic and representative of experiment?
Although some parameters can be chosen roughly according to common chemical intuition, for other quantities, the typical behaviour is unclear. One such quantity is the electronic transition dipole moment $\mu_{eg}(x)$; how it varies along a normal mode $x$ determines by how much the London dispersion interaction changes when exciting this normal mode. Future efforts should be directed into mapping out this quantity \emph{ab initio} for a set of molecules and modes relevant to experiment, to get a sense of how strongly the dispersion interaction typically depends on vibrational state.   

Another important open question pertains to the accuracy of the approach we used to run dynamics on a model system, from which we extracted cavity-modified rate constants. 
In our mixed quantum--classical dynamics scheme, we have neglected any back-reaction that the classical system (the reaction coordinate) might have on the quantum system (the perpendicular vibrational mode, the cavity mode, and all other molecular modes coupled to the cavity) (note that what we call back-reaction is the diametric opposite of what is meant by back-reaction in nonadiabatic dynamics, namely the force the electrons exert onto the nuclei). One might think this should be straightforward to include, as standard mixed quantum--classical methods, such as surface hopping and Ehrenfest dynamics, do account for it, at least to some level of approximation.
Note however that, unlike in standard nonadiabatic dynamics, transitions in the quantum part of the system are caused by external baths (coupling to the solvent and to the electromagnetic modes outside the cavity, described by phenomenological vibrational relaxation and cavity loss rates), rather than by the combination of motion of the classical coordinate and nonadiabatic couplings between the quantum states.

The lack of back-reaction in our scheme means that, while the state of the quantum system affects the energy gradient that the classical system feels, the opposite is not true: we have assumed the hopping rates between quantum states to be independent of the position and velocity of the classical trajectory. Including back-reaction from the classical system onto the quantum system may be important, but it is not trivial to account for; it would likely require one to abandon the phenomenological decay rates we have been working with, and instead explicitly use harmonic baths with well-defined spectral densities. One will need to derive what the back-reaction looks like, and then devise an appropriate mixed quantum--classical dynamics scheme for it.

Alternatively, one could describe the whole system quantum-mechanically, and run quantum dynamics simulations. This is certainly more reliable, and would additionally provide a way to answer another major open question: does the cavity-modification of the reaction rate we found for two molecules persist as we add more molecules to the cavity?
Abandoning the mixed quantum--classical scheme and treating the full system quantum-mechanically instead would relieve us from the burden of the secular approximation -- it would open the way to including higher excitation manifold, and thereby converging the calculation for more than two or three molecules.



Next to the investigations into the foundations of our approach described above, many other exciting directions for further research lay open. For example, in our calculation of the dispersion interaction, we have relied on second-order perturbation theory and truncated the multipole expansion of the Coulomb interaction at lowest order (yielding the dipole-dipole interaction). We have also assumed the exchange interaction between molecules $A$ and $B$ to be negligible. This is valid as long as the molecules are far apart. Bringing them closer together requires one to go beyond the dipole approximation, and retain higher orders of the multipole expansion (or avoid using the multipole expansion altogether) \cite{stone2013theory}. One might also need to go to higher order in perturbation theory; this would moreover open the way to studying three-body (Axilrod--Teller--Muto) dispersion interaction terms \cite{stone2013theory}. Both of these directions can potentially unveil more exciting effects of the cavity on intermolecular interactions. 

Ultimately, a non-perturbative approach towards calculating vibrationally-resolved London dispersion forces would be desirable -- this could involve working in a nuclear-electronic orbital framework \cite{pavosevic2020NEO}, which accounts for the quantum nature of certain nuclei while solving the electronic structure problem. Treating both molecules $A$ and $B$ together in an \emph{ab initio} calculation also accounts for exchange interactions between them, which become important as the molecules come very close to each other.

Other interesting lines of enquiry include incorporating levels of complication that have been left out of our analysis. For example, we have treated the molecular vibrations as harmonic oscillators; it would be interesting to see whether incorporating anharmonicity leads to any significant changes. 
Also, in our model we kept the molecular frequencies and orientations fixed over the course of the reaction. In reality, however, the timescale of the reaction much longer than the timescale associated with changes in the molecular orientation or in its environment, affecting the vibrational frequency. These may therefore change over the course of a reaction, thereby affecting the dispersion forces that influence the reaction. Another effect we left out from our description is the coupling \emph{between} vibrational states that arises from virtual excitations to high-lying electronic excited states. Using an approach very similar to Eq.~\eqref{eq:vibdisp}, one can construct the entire effective low-energy Hamiltonian \cite{landi2024eigenoperator}, which features the dispersive shifts of Eq.~\eqref{eq:vibdisp} on the diagonal. The new, off-diagonal couplings could facilitate another way to hop between states, and thereby in principle affect the dynamics.

All in all, we hope that by offering a fresh view, this work spurs a new line of investigation, and perhaps moves us a step closer towards understanding VSC-modified chemistry.

\section{Methods}
\subsection{Vibrationally-resolved London dispersion interaction}
The expression for the London dispersion interaction between molecules $A$ and $B$ outside the cavity is given in Eq.~\eqref{eq:C6nocav}, as a function of the vibrational states of both of the molecules. To evaluate the expression, we need the functional form of the electronic transition dipole moment, and the wavefunctions of the relevant vibrational states. Throughout this work, we have assumed the vibrational modes to be harmonic, meaning we have analytical forms of the wavefunction $\psi_\alpha(x)$ for each vibrational state $\alpha$ at our disposal. We set the frequency to $\omega_g=$ \SI{550}{\per\centi\meter}, or a value close to this in case of frequency disorder (see below). We also require the electronic transition dipole moment $\mu_{eg}(x)$ to be a function of the nuclear coordinate $x$, because if it were a constant, this would lead to all vibrational states having exactly the same $C_6$ coefficients, removing all interesting effects. For all results shown in the main text, we have taken the transition dipole moment as   
\begin{equation}\label{eq:mu_egfunc}
    \mu_{eg}(x)=7\times 10^{-5}(120+x)^2+0.018x 
\end{equation}
where $x$ is in mass-weighted coordinates, and the result is in a.u. 
The ``square-dipole" matrix elements that we need are given by the integral
\begin{equation} \label{eq:Mab_integral}
\begin{aligned}
       \mathrm{M}_{\alpha\beta} &= \bra{\alpha} \mu_{eg}(\hat{x})^2 \ket{\beta} \\
       &=\int \dd x \, \psi_\alpha^*(x) \mu_{eg}(x)^2 \psi_\beta(x)
\end{aligned}
\end{equation}
which we evaluate numerically using adaptive quadrature.

The functional form in Eq.~\eqref{eq:mu_egfunc} leads to a transition dipole moment of a few Debye when the molecule is its vibrational ground state, and a slightly larger value for the vibrational excited state -- both of these are fairly typical values for a transition dipole moment. Also, for this choice $\mathrm{M}_{01}$ is only a little smaller than $\mathrm{M}_{00}$ and $\mathrm{M}_{11}$, which becomes important when discussing the influence of a cavity on these dispersive shifts:
in this case, the polariton $C_6$ coefficients are larger than that of the bare excited state (\emph{cf.} Eq.~\eqref{eq:C6polariton}). In the SI, we present results for a linear form of $\mu_{eg}(x)$, with parameters tuned to produce qualitatively different results in this respect. It leads to similar results for the $N=2$ rate enhancement, and the collective  suppression of the variance of $C_6$.

\subsection{VSC effect on dispersion interactions}
Coupling to a cavity mode affects the eigenstates of the system, which in turn leads to altered dispersion interactions. We describe the cavity mode and the $N$ vibrational modes of the molecules as a system of coupled harmonic oscillators in the rotating wave approximation (RWA), 
\begin{equation} \label{eq:H}
    \hat{H}=\sum_{i=1}^{N}\hbar \omega_i \big(\hat{a}_i^\dagger \hat{a}_i + \tfrac{1}{2}\big) + \hbar \omega_\mathrm{c} \big(\hat{a}_\mathrm{c}^\dagger \hat{a}_\mathrm{c} + \tfrac{1}{2}\big) + \hbar  \sum_{n=1}^N g_i(\hat{a}_\mathrm{c}^\dagger \hat{a}_i + \hat{a}_\mathrm{c} \hat{a}_i^\dagger).
\end{equation}
Here, $\hat{a}_i$ ($\hat{a}_i^\dagger$) is the annihilation (creation) operator of the vibrational mode of molecule $i$, $\omega_i$ the corresponding frequency (close to \SI{550}{\per\centi\meter}, \emph{vide infra}), $\hat{a}_\mathrm{c}$ ($\hat{a}_\mathrm{c}^\dagger$) is the annihilation (creation) operator of the cavity mode, and $\omega_\mathrm{c}$ is the cavity frequency. Finally, $g_i$ is the light--matter coupling strength between the cavity mode and the $i$th molecule. 

Note that, when truncating to only the ground state and single-excitation subspace, this Hamiltonian is equivalent to the Tavis--Cummings model, of which we used the analytical eigenstates in Sec.~\ref{sec:LD_VSC}.
It should also be remarked that due to the RWA, there is no coupling between the excitation manifolds, and as a consequence of that, the ground state remains unaffected by the cavity coupling. 

We have not included the dipole self-energy (DSE) term in Eq.~\eqref{eq:H}, as our observable here measures changes in the wavefunction, rather than the ground-state energy. For small $g_\mathrm{c}$, as used here, including it would mainly result in a slight frequency shift in the results. By neglecting the DSE altogether we also avoid the question of whether or not to include DSE cross terms in the Hamiltonian \cite{fiechter2025hamiltonian}.

To obtain the eigenstates of $\hat{H}$ in the case of disorder, we construct it in the uncoupled basis, where we can express eigenstates as $\ket{k} = \ket{n_A,n_B, ..., n_\mathrm{c}}$ with $n_i$ the number of vibrational excitations in mode $i$, and $n_\mathrm{c}$ the number of excitations in the cavity mode. We truncate the basis based on the total number of excitations per state, keeping only the states that have fewer excitations than a specified maximum. We numerically diagonalize the resulting matrix  and obtain the new eigenstates $\ket{j}$ in terms of the original ones, $\ket{j}=\sum_k U_{jk}\ket{k}$. Now, the $C_6$ coefficients in the new basis are also straightforward to evaluate:
\begin{equation} \label{eq:C6general}
C_6(\ket{j} ) =  \frac{1}{2\hbar \omega_{eg}}  \sum_{k,\ell} U^*_{jk} U_{j\ell} \bra{k} \mu_{eg}(\hat{x}_A)^2\mu_{eg}(\hat{x}_B)^2 \ket{\ell}.
\end{equation}
The ``squared-dipole" matrix elements in this equation are now expressed in terms of the original excitation number basis, where they are straightforward to evaluate (Eq.~\eqref{eq:Mab_integral}).

\subsection{Disorder in the frequencies and coupling strengths}
We incorporate disorder in the vibrational frequencies of the molecules by sampling them from a Gaussian distribution with  $\sigma=\SI{5}{\per\centi\meter}$, centered at \SI{550}{\per\centi\meter}. For the $N=2$ results in the main text, this resulted in $\omega_i=$ \SI{549.97}{\per\centi\meter} for molecule $A$ and $\omega_i=$ \SI{555.23}{\per\centi\meter} for molecule $B$. When adding more molecules (as in Figs.~\ref{fig:C6_8mol}, \ref{fig:C6_manymol} and \ref{fig:Ndependence}), we sample their frequency and add it to the list (rather than resampling all frequencies). The frequencies of the 20 molecules shown in Fig.~\ref{fig:C6_manymol} are given in the SI. 

In Figs.~\ref{fig:C6_manymol} and \ref{fig:Ndependence}, we have also accounted for orientational and positional disorder. Both of these affect the light--matter coupling strength $g_i$. Orientational disorder leads to imperfect alignment of the molecules' ground-state dipole moments $\boldsymbol{\mu}_i$ with the cavity mode. This diminishes $g_i$ as in the full (Pauli--Fierz) Hamiltonian, the coupling strength is proportional to $\boldsymbol{\mu}_i\cdot \mathrm{e}_\lambda$, where $\mathrm{e}_\lambda$ is the polarization direction of the cavity mode. Assuming an isotropic distribution of orientations, we account for this by multiplying $g_\mathrm{c}$ by the $z$-component of a randomly oriented unit vector $\hat{u}_z$ for each molecule $i$.

The light--matter coupling strength is also affected by the molecule's position in the cavity: cavity modes are standing waves, so that their electric field amplitude is modulated as a function of position along the cavity axis 
\cite{scully1999quantum}. We assume coupling to the first cavity mode, assume a constant concentration (uniform spatial distribution of the molecules), and set the spacer width to be a tenth of the cavity length -- so that the molecules cannot come closer to the mirrors than 10\% of the cavity length. 
Overall, this yields 
\begin{equation}
    g_i = g_\mathrm{c} \,\hat{u}_z \, \sin(\pi \times a)
\end{equation}
where $a$ is a uniformly-distributed random number between 0.1 and 0.9, and $g_\mathrm{c}$ is the maximal coupling strength, taken as an input parameter. 

Note that in all of the above, we keep $A$ and $B$ fixed at an antinode, and aligned to cavity mode polarization direction, so that their $g_i=g_\mathrm{c}$. The rescaling factors $g_i/g_\mathrm{c}$ for all other molecules used in Fig.~\ref{fig:C6_manymol} can be found in the SI. 

\subsection{Dynamics and rate constant calculations}

In Sec.~\ref{sec:reactionN2}, we introduced a simple model for describing the effect of VSC on a reaction in which the dispersion interaction plays an important role. Here, we elaborate on the mixed quantum--classical approach we used to compute the rate constant for this system.

We divide the system into two parts: the reaction coordinate of molecule $A$, which we treat classically; and the perpendicular vibrational modes of all molecules as well as the cavity mode, which we treat quantum-mechanically. 

We start by diagonalizing the Hamiltonian for the quantum part of the system (Eq.~\eqref{eq:H}). The population dynamics between the eigenstates can be described with a Pauli master equation \cite{breuer2002theory,may2023charge, du2022catalysis}, provided that the eigenstates are well-separated in energy compared to their linewidth \cite{du2022catalysis}, so that the secular approximation is satisfied. This can be achieved by choosing the vibrational frequencies of the molecules different enough ($\gtrsim $ \SI{1}{\per\centi\meter}), and truncating to the first excitation manifold, where the states do not cross for any $\omega_\mathrm{c}$. In higher excitation manifolds, which need to be included as we increase the number of molecules, the states do cross, leading to a breakdown of the secular approximation for almost all frequencies around resonance; see also the SI. This therefore falls outside the range of validity of our propagation scheme.  

As our quantum system is just an ensemble of coupled harmonic oscillators, it is straightforward to derive such a population-dynamics equation from the Lindblad equation \cite{carmichael1999statistical,meystre2007elements}. The derivation for the general case is given in the SI; when restricting to $N=2$ and the first excitation manifold, our Pauli master equation reads
\begin{equation} \label{eq:pjt}
    \frac{\dd p_j}{\dd t} = \sum_{k\neq j}w_{j\leftarrow k}p_k - p_j \sum_{k\neq j} w_{k\leftarrow j}
\end{equation}
where 
\begin{equation}
    \mathbf{w} = \begin{pmatrix}
        0 & \tilde{\gamma}_\mathrm{LP}^\downarrow & \tilde{\gamma}_\mathrm{D}^\downarrow  & \tilde{\gamma}_\mathrm{UP}^\downarrow \\
        \tilde{\gamma}_\mathrm{LP}^\uparrow & 0 & 0 & 0\\
        \tilde{\gamma}_\mathrm{D}^\uparrow & 0 & 0 & 0\\
        \tilde{\gamma}_\mathrm{UP}^\uparrow & 0 & 0 & 0\\
    \end{pmatrix}.
\end{equation}
The decay rate of eigenstate $q=\mathrm{LP}, \mathrm{D}, \mathrm{UP}$ can be expressed in terms of the phenomenological cavity loss rate $\kappa$ and vibrational relaxation rate $\gamma_\mathrm{vib}$. Summarizing these original rates as $\gamma_i$, with $\gamma_i=\kappa$ for $i=0$ (cavity mode), and $\gamma_i=\gamma_\mathrm{vib}$ otherwise (vibrational mode), we obtain the decay rates
\begin{equation}
    \tilde{\gamma}_q^{\downarrow}=\sum_i|c_{iq}|^2 \gamma_i (\bar{n}(\omega_i)+1).
\end{equation}
Here, $c_{iq}$ is the coefficient relating the eigenmodes to the original modes; it can be obtained by diagonalizing Eq.~\eqref{eq:H} for the first excitation manifold only. $\bar{n}(\omega_i)$ is the thermal occupation number of bare mode $i$, 
\begin{equation}
    \bar{n}(\omega_i)=\frac{\mathrm{e}^{-\beta\hbar\omega_i}}{1-\mathrm{e}^{-\beta\hbar\omega_i}}.
\end{equation}
The upwards rates are related to the decay rates by detailed balance,
\begin{equation}
    \tilde{\gamma}_q^{\uparrow}= \mathrm{e}^{-\beta\hbar\omega_q}\tilde{\gamma}_q^{\downarrow} .
\end{equation}

The bare cavity decay rate $\kappa$ is related to the cavity $Q$-factor by $\kappa = \omega_\mathrm{c}/Q$ \cite{fox2006quantum}; for a cavity mode tuned to \SI{550}{\per\centi\meter} and $Q=86.2$, this corresponds to a lifetime at room temperature of $1/(\kappa(\bar{n}(\omega_\mathrm{c})+1)) \approx $ \SI{0.8}{\pico\second}. Typical vibrational lifetimes are much longer than that; we set the lifetime at room temperature to $1/(\gamma_\mathrm{vib}(\bar{n}(\omega_i) +1)) = $  \SI{50}{\pico\second}. This is similar to Refs.~\cite{du2022catalysis,xiang2019polariton}, albeit slightly lower, as our vibration is at a lower frequency.

In the above, we have only included transitions between the ground state and the first excitation manifold. Accounting for dephasing results in transitions within the first excitation manifold, \emph{i.e.} between dark states and polaritons, filling up the lower right block in the $\mathbf{w}$-matrix too. Dephasing rates are calculated as discussed in the Supplemental Material of Ref.~\citenum{du2022catalysis}, for the same set of bath parameters. As shown in our SI, we find that including dephasing has a negligible effect on the reaction rates we calculate. 

We generate stochastic trajectories representative of Eq.~\eqref{eq:pjt} using a kinetic Monte Carlo (kMC) algorithm \cite{peters2017reaction}. This gives us the state that the quantum system is in, and for how long it remains in this state. It is this information that is passed on to the classical part of the system: in surface-hopping terminology, it sets the ``active surface", \emph{i.e.} which potential energy surface the classical trajectory feels. 

For all states, the potential energy surface is given by an asymmetric double-well potential of the form
\begin{equation}
    V_j(x) = V_j^\ddagger \bigg(\frac{x^2}{x_0^2} -1\bigg)^2 - c x .
\end{equation}
The asymmetry parameter $c$ is kept constant at $c=1\times 10^{-5}$ a.u., and $x_0=85$ a.u. (note that we work in mass-weighted coordinates). We let the barrier height parameter $V^\ddagger$ depend on the $C_6$ coefficient of the active state:
\begin{equation}
    V^{\ddagger}_j = V^{\ddagger}_{\mathrm{G}} - V' f(C_{6,j} / C_{6,\mathrm{G}}).
\end{equation}
We set the ground state barrier height parameter $V^{\ddagger}_{\mathrm{G}}=$ \SI{1500}{\per\centi\meter} (as a side note, this is only equal to the actual energy difference between minimum and transition state if we set $c=0$). We let the barrier height parameter for the $j$th state depend linearly on the ratio of its $C_6$ coefficient compared to that of the ground state:
\begin{equation}
    f(C_{6,j}/C_{6,\mathrm{G}}) = C_{6,j}/C_{6,\mathrm{G}} -1.
\end{equation}
The strenght of this dependence is determined by the parameter $V'$, which we set to \SI{1000}{\per\centi\meter}.

We start the simulation by randomly picking a quantum state $j$ with a probability given by its Boltzmann weight. We then initialize the initial position of the classical coordinate by sampling the corresponding distribution, $\mathrm{e}^{-\beta V_j(x)}$, in the reactant well (left). We propagate the trajectory in time according to the Langevin equation with friction $\xi$ and on potential energy surface $V_j$. We do this for as long as the kMC algorithm says we are in state $j$; after this, the trajectory hops to the next state it prescribes. 

We propagate 18'000 trajectories, each for $7\times10^6$ a.u., with a 10 a.u.~time step. For each trajectory, we track whether it is on the reactant (left) or the product (right) side of the barrier. Averaging over many trajectories gives us the product population $P_\mathrm{P}(t)$ as a function of time. After some initial transient dynamics, the relaxation of $P_\mathrm{P}(t)$ to equilibrium is exponential in behaviour (see SI), as expected from a first-order rate process. We extract the forward and backward rate constants by fitting our results to \cite{peters2017reaction}
\begin{equation}
    P_\mathrm{P}(t) = A \mathrm{e}^{-k_\mathrm{eff}t} + P_\mathrm{P, eq}.
\end{equation}
where $k_\mathrm{eff} = k_{\mathrm{R}\rightarrow\mathrm{P}} + k_{\mathrm{R}\rightarrow\mathrm{P}}$.
Using that $P_\mathrm{P, eq}/P_\mathrm{R, eq} = k_{\mathrm{R}\rightarrow\mathrm{P}}/k_{\mathrm{P}\rightarrow\mathrm{R}} $ by detailed balance, we can extract the forward reaction rate from the fitting parameters as $k_{\mathrm{R}\rightarrow\mathrm{P}} = k_\mathrm{eff} P_\mathrm{P,eq}$. The error bars shown in the plots are the 68\% bootstrap confidence intervals. We elaborate on this in the SI.


\subsection{Distribution of population over the excitation manifolds}
In Fig.~\ref{fig:exman_population}, we plotted the population in each of the excitation manifolds for varying numbers of molecules $N$, assuming that all molecules and the cavity mode have the same frequency ($\omega = $ \SI{550}{\per\centi\meter}) and are uncoupled.

The population of the $n$th excitation manifold is given by $g_n \mathrm{e}^{-\beta E_n}/Z$. Here $E_n=n\hbar\omega$; $Z$ is the total partition function, which is simply the product of the partition functions $Z_\mathrm{HO}(\beta,\omega)$ of each of the uncoupled harmonic oscillators: $Z=Z_\mathrm{HO}(\beta,\omega)^{N+1}$. 

The degeneracy factor $g_n$ is equal to the number of states with $n$ excitations. Calculating it is boils down to a stars-and-bars problem in combinatorics, \emph{i.e.} 
\begin{equation}
\begin{aligned}
        g_n &= \binom{n+(N+1) -1}{(N+1)-1} \\
        &= \binom{n+N}{N}\\
        &=\frac{(n+N)!}{N!\,n!}.
\end{aligned}
\end{equation}

\vspace{-0.2cm} 

\bibliography{ref}

\end{document}